\documentclass[prl,amsmath,amssymb,twocolumn,superscriptaddress]{revtex4-1}
\usepackage{graphicx,bm,color,xcolor,appendix}
\usepackage{ulem}
\usepackage[colorlinks,bookmarks=true,citecolor=blue,linkcolor=blue,urlcolor=blue]{hyperref}

\newcommand{\mr}{moir\'e~}
\newcommand{\Mr}{Moir\'e~}
\newcommand{\wsetwo}{WSe$_2$~}
\DeclareMathOperator{\sign}{sgn}

\newcommand{\mC}{\mathcal{C}}

\newcommand{\mT}{\mathcal{T}}

\begin{document}
\title{Staggered Pseudo Magnetic Field in Twisted Transition Metal Dichalcogenides: Physical Origin and Experimental Consequences}

\author{Jie Wang}
\affiliation{Center for Computational Quantum Physics, Flatiron Institute, 162 5th Avenue, New York, New York 10010, USA}
\author{Jiawei Zang}
\affiliation{Department of Physics, Columbia University, 538 W 120th Street, New York, New York 10027, USA}
\author{Jennifer Cano}
\affiliation{Center for Computational Quantum Physics, Flatiron Institute, 162 5th Avenue, New York, New York 10010, USA}
\affiliation{Department of Physics and Astronomy, Stony Brook University, Stony Brook, New York 11974, USA}
\author{Andrew J. Millis}
\affiliation{Center for Computational Quantum Physics, Flatiron Institute, 162 5th Avenue, New York, New York 10010, USA}
\affiliation{Department of Physics, Columbia University, 538 W 120th Street, New York, New York 10027, USA}

\begin{abstract}
Strong magnetic fields profoundly affect the quantum physics of charged particles, as seen for example by the integer and fractionally quantized Hall effects, and the fractal `Hofstadter butterfly' spectrum of electrons in the presence of a periodic potential and a magnetic field. Intrinsic physics can lead to effects equivalent to those produced by an externally applied magnetic field. Examples include the `staggered flux' phases emerging in some theories of quantum spin liquids and the Chern insulator behavior of twisted bilayer graphene when valley symmetry is broken. In this paper we show that when two layers of the transition metal dichalcogenide material \wsetwo are stacked at a small relative twist angle to form a \mr bilayer, the resulting low energy physics can be understood in terms of electrons moving in a strong and tunable staggered flux. We predict experimental consequences including sign reversals of the Hall coefficient on application of an interlayer potential and spin currents appearing at sample edges and interfaces.
\end{abstract}

\maketitle

\Mr bilayers are formed when two atomically thin layers are stacked at a small relative twist angle. The band properties of \mr bilayers are easily tuned by changes in twist angle, stacking and gate voltage, making \mr materials a versatile platform for studying many aspects of electronic physics \cite{Bistritzer12233,Cao:2018aa,Cao:2018ab,Sharpe605,Serlin900,Stepanov:2020aa,Tang:2020aa,Wang:2020aa,Li:2021aa,Ghiotto:2021aa,Jin:2021aa,Andersen:2021aa,Xu:2020aa,FengchengWu_PRL18,FengchengWu_PRL19,wang2021onedimensional,Liang_moireTMD21,Liang_spintexture21,Macdonald_TMD21,Macdonald_moireHubbard21,Liang_SC_tTMD21,DasSarma_Int_Range21,yahui_electricaldetectionsplinliquid,Kennes_tTMD_spinliquid}.

The low energy physics of \mr \wsetwo is well captured by the \mr Hubbard model $H=H_0+H_I$ describing interacting electrons hopping  on a \mr triangular lattice \cite{Tang:2020aa,Wang:2020aa,Li:2021aa,Ghiotto:2021aa,DasSarma_tTMD_PRR,Jiawei_HartreeFock}. The interaction part $H_I$ is normally taken to be of the Hubbard type $\sum_{\bm r}Un_{\bm r\uparrow}n_{\bm r\downarrow}$ with $U>0$ while the kinetic part of $H$ is
\begin{eqnarray}
H_0 &=& -\sum_{\bm r, \bm r',\sigma}t_{\bm r,\bm r'}e^{i\phi_{\bm r,\bm r',\sigma}}c^{\dag}_{\bm r',\sigma}c_{\bm r,\sigma}.\label{moirehubbard}
\end{eqnarray}

\begin{figure}[t]
\centering
\includegraphics[width=1.0\linewidth]{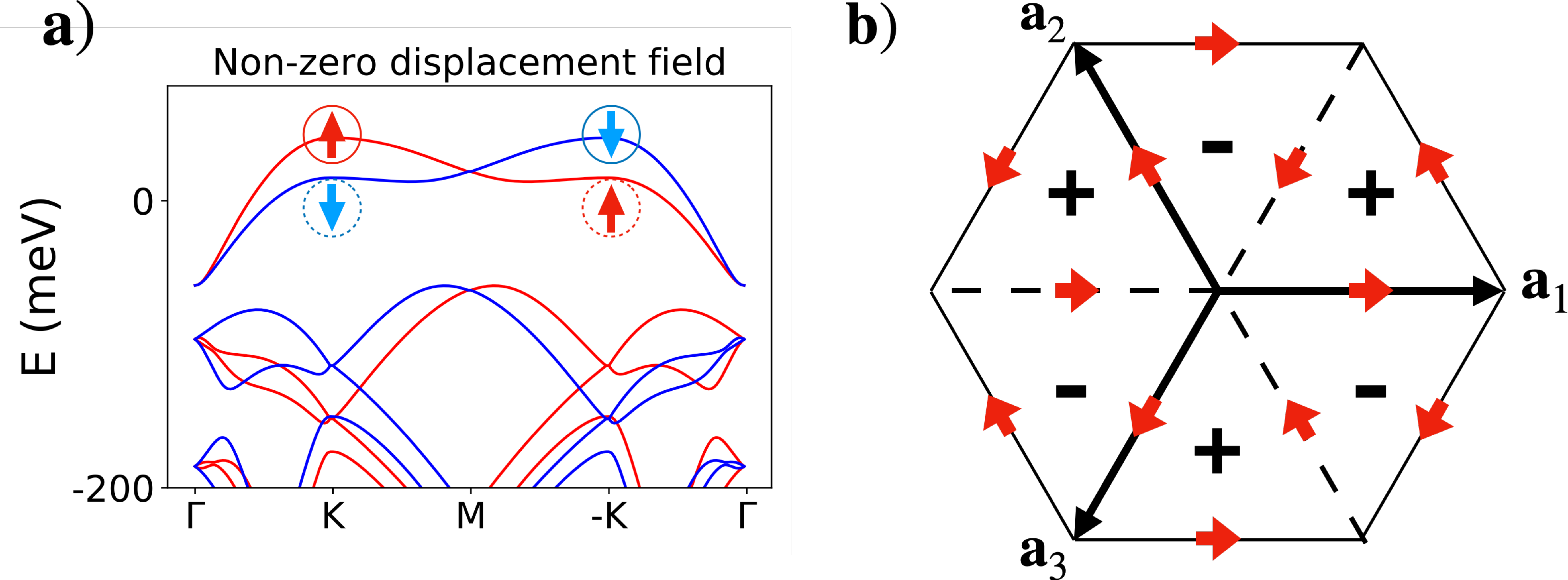}
\caption{(a) Sketch of the band structure in twisted \wsetwo shows spin split bands in the presence of a nonzero displacement field, where the red (blue) indicates the bands from the $\bm K_0$ ($\bm K^{\prime}_0$) valley of the monolayer TMD, and the solid (empty) circle indicates the bands from the top (bottom) layer. (b) Representation of the hopping amplitude and phase. The vertices correspond to triangular lattice sites of the \mr lattice, where $\bm a_{1,2,3}$ are vectors to nearest neighbors. The sign of the spin up hopping phases are represented by the red arrows, which give rise to the staggered effective magnetic fields whose signs are marked by $\pm$ at the center of triangles.}\label{fig:band}
\end{figure}

Here $\bm r$ and $\bm r'$ label sites of a triangular lattice, $c^{\dag}_{\bm r,\sigma}$ creates an electron of spin $\sigma=\uparrow,\downarrow$ on site $\bm r$, and $t_{\bm r,\bm r'}$ is a positive number giving the modulus of the hopping amplitude between sites $\bm{r}$ and $\bm{r^\prime}$. The crucial new feature of the \mr Hubbard model is the spin-dependent phase $\phi_{\bm r,\bm r',\sigma}$ appearing in the hopping. This phase arises from the interplay of the \mr structure and the strong spin orbit coupling of the individual layers, affects the band structure as shown in Fig.~\ref{fig:band} (a), and is experimentally tunable by varying the ``displacement field", i.e., the voltage difference between layers.

Previous work has focused on the role of the phase in tuning the energy, momentum-space position and nature of the van Hove singularities in the dispersion \cite{Bi:2021aa,DasSarma_tTMD_PRR,Jiawei_HartreeFock}. However, the phase factor may be viewed as the gauge field arising from a spin-dependent staggered magnetic flux which alternates between elementary triangles [see Fig.~\ref{fig:band} (b)]; in other words, the \mr Hubbard model is properly thought of as a model of electrons moving in a strong and tunable staggered flux. Here we show that the staggered field has important observable consequences: the Hofstadter butterfly spectrum that emerges when a uniform magnetic field is applied acquires a nontrivial structure that implies tunable sign reversals of the Hall conductivity, while spatial gradients of the displacement field or an interface between two different values of the displacement field produce spin currents at edges or interfaces.

\textbf{Origin of the spin dependent staggered magnetic field.}
While the staggered flux is revealed in band theory calculations, it is worth discussing it from a general symmetry point of view. In monolayer form, a TMD material consists of a triangular lattice of transition metal ions sandwiched between two triangular lattices of chalcogen atoms. The band structure has two valleys \cite{Manzeli:2017aa}. In WSe$_2$, the valence band maxima occur at the Dirac points $\bm K_0$ and $\bm K_0^\prime$ of the monolayer Brillouin zone. Due to the strong spin orbit coupling, spin is quantized perpendicular to the plane, such that states near $\bm K_0$ have spin up while those near $\bm K_0^\prime$ have spin down. The two valleys are related by time reversal.

Stacking two layers of TMD with a relative twist angle between them creates an enlarged \mr triangular lattice and enriches the system by an intricate interplay between spin, valley and layer degrees of freedom. The low-energy \mr band structure is well described by a tight-binding model on the \mr triangular lattice \cite{DasSarma_tTMD_PRR,Jiawei_HartreeFock}. An applied displacement field preserves the \mr translation symmetry, the three-fold rotation symmetry $\mC_3$ of the \mr lattice, and time-reversal symmetry $\mT$.
From this symmetry point of view, the only place that the displacement field can affect the electron's dynamics is through modifying the hopping term by a spin dependent phase factor $\exp(i\phi_{\bm r,\bm r',\sigma})$ as shown in Eqn.~(\ref{moirehubbard}) \footnote{The displacement field also affects the magnitude of the hopping but this effect is not directly relevant for our calculations and is not notated here.}. The concrete dependence of this term on the displacement field can be obtained from first principle calculations \cite{Wang:2020aa}. Due to the $\mC_3$ symmetry and the time reversal symmetry $\mT$, the phase fields are constrained to the following form:
\begin{equation}
    \phi_{\bm r,\bm r+\bm a_n,\sigma} = \sigma\phi,\quad n=1,2,3,\label{defphi}
\end{equation}
where $\sigma=+1$ for spin up,  $-1$ for spin down, and the combination of $C_3$ symmetry and interlayer inversion means that $\phi=0$ at zero displacement field. Here $\bm a_{1,2}$ are the two independent Bravais lattice vectors of the triangular lattice, and $\bm a_3=-(\bm a_1+\bm a_2)$ [see Fig.~\ref{fig:band} (b)]. For simplicity we only retained the nearest neighbor (NN) hopping term. Then, $H_0$ of the \mr Hubbard model is simplified to $H_0=H_{0}^{\uparrow}+H_{0}^{\downarrow}$, where the spin up part is:
\begin{equation}
H_{0}^{\uparrow} = -t\sum_{\bm r}\sum_{n=1,2,3}e^{i\phi}c^{\dag}_{\bm r+\bm a_n,\uparrow}c_{\bm r,\uparrow} + H.c,\label{moirehubbard2}
\end{equation}
and the spin down part is obtained by the time-reversal symmetry operation.

Standard gauge invariance arguments show that the net phase accumulated around the triangular plaquette $\bm r\rightarrow\bm r+\bm a_1\rightarrow\bm r-\bm a_3\rightarrow\bm r$ is $3\sigma\phi$, and that accumulated on the adjacent plaquette is $-3\sigma\phi$. Thus, for a uniform displacement field, the electrons feel a staggered magnetic field, which alternates in sign between adjacent triangles so the net effective magnetic field is averaged to zero [see Fig.~\ref{fig:band}(b)]. We now consider some physical consequences of the staggered field.

\textbf{Displacement field tunable Hofstadter butterfly.}
The combination of a uniform magnetic field and a periodic potential leads to a self-similar recursive spectrum, known as the ``Hofstadter butterfly'' \cite{Hofstadter_butterfly}. The Hofstadter butterfly is modified by a staggered magnetic field \cite{Li_2011}. The modifications are independent of the sign of the staggered field, so the spectrum for spin up and spin down electrons will be the same.

We have computed the Hofstadter butterfly electronic spectrum following from Eqn.~(\ref{moirehubbard2}). We present our results in terms of two parameters: $\phi$, the hopping phase induced by the displacement field defined in Eqn.~(\ref{defphi}) and $\Phi_B=2\pi p/q$, the flux per \mr unit cell (consisting of two adjacent elementary triangles) arising from the uniform applied magnetic field. We observe that the spectrum  is periodic in $\phi\rightarrow\phi\pm2\pi/3$, which is a result of the fact that inserting staggered $\pm2\pi$ flux can be trivially removed by a gauge transformation \cite{Jiawei_HartreeFock}. We further note that the spectrum found for $(\phi,\Phi_B)$ is identical to that found for $(\phi\pm\pi/3,-\Phi_B)$, because a particle-hole transformation maps $\phi$ to $\phi+\pi/3$ and $\Phi_B$ to $-\Phi_B$. Last but not least, the butterfly spectrum is invariant under a change in sign of the staggered flux $\phi\rightarrow-\phi$. Following the standard treatment of the Hofstadter problem \cite{Li_2011}, the model parameterized by $(\phi,\Phi_B)$ can be straightforwardly diagonalized. The symmetry considerations mean that it is sufficient to look at the energy spectrum for $\phi\in[0,\pi/6]$. 

\begin{figure*}[ht]
\centering
\includegraphics[width=0.75\linewidth]{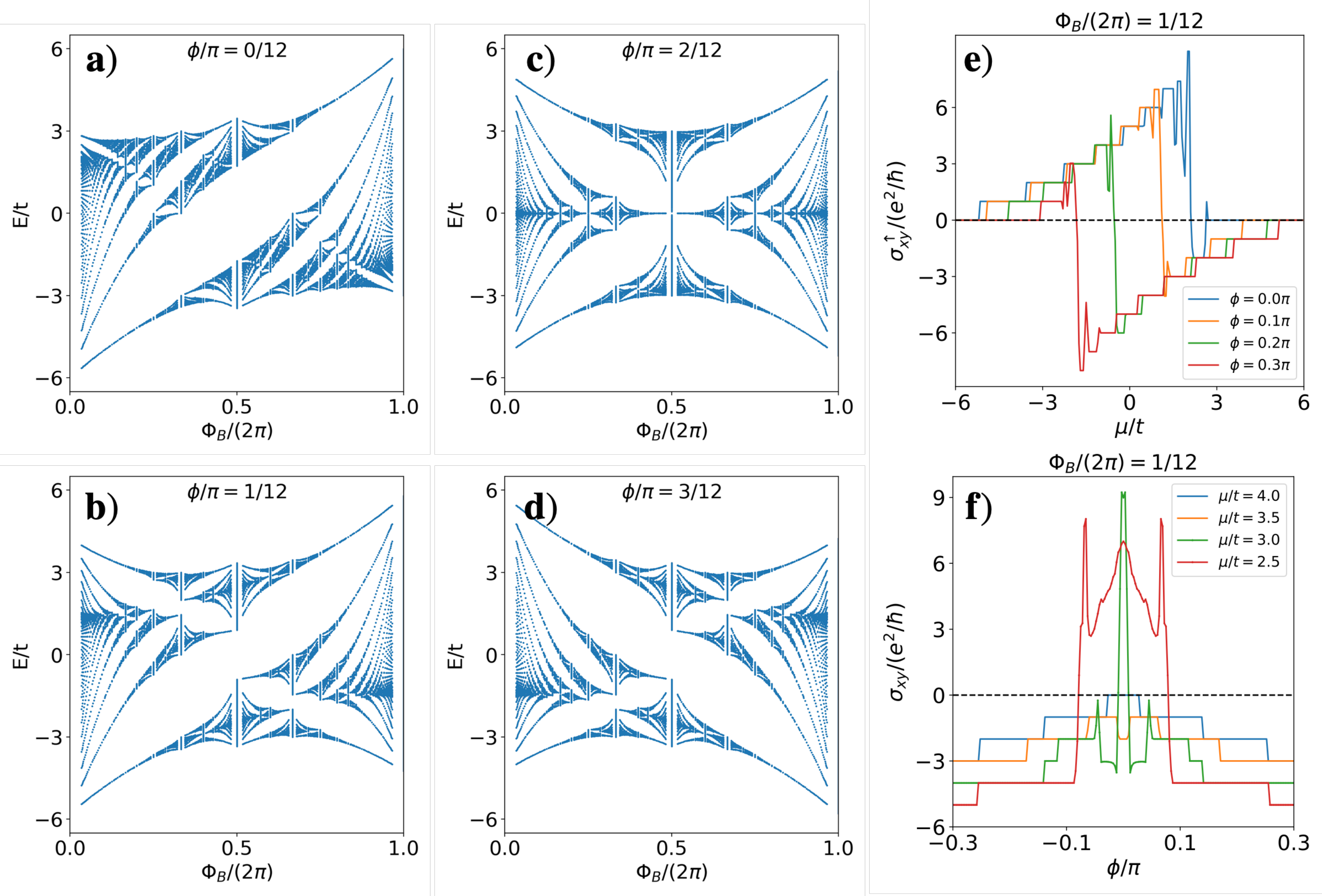}
\caption{(a) to (c): Evolution of the Hofstadter butterfly in the presence of a uniform magnetic field $\Phi_B$ for different values of the displacement field induced phase $\phi\in[0,\pi/6]$. (d) shows the Hofstadter butterfly at $\phi/\pi = 3/12$ and is related to (b) under inverting the sign of $\Phi_{B}$ due to the symmetries discussed in the main text. (e) and (f): The Hall conductance as a function of chemical potential or the phase $\phi$ shows a change of the sign of Hall conductance, estimated at twist angle $\theta\sim3^{\circ}$ and $10$T external magnetic field.}\label{fig:ubutterfly}
\end{figure*}

Our calculated Hofstadter butterfly spectra are shown in Fig.~\ref{fig:ubutterfly} from (a) to (d) as a function of uniform applied field $\Phi_B$ at several different values of $\phi$. The entire electronic spectrum is seen to be strongly tunable with displacement field. The quantization of the spectrum into isolated Landau levels is visible, along with breaking and reconnection of the subbands in a manner that depends strongly on displacement field and applied magnetic field. Breaking and reconnection of Landau subbands is known to be associated with changes in sign of the Hall conductance~\cite{Li_2011}. Panel (e) shows the Hall conductivity of the spin up branch computed from the standard Kubo formula \cite{kubo1,kubo2,TKNN}:
\begin{equation}
    \frac{\sigma^{\uparrow}_{xy}(\mu)}{e^2/\hbar} = \sum_{\epsilon^{\uparrow}_m<\mu<\epsilon^{\uparrow}_n}\int\frac{d^2\bm k}{(2\pi)^2}\frac{i\epsilon_{ab}(J^a_{\bm k})_{mn}(J^b_{\bm k})_{nm}}{(\epsilon_m^\uparrow-\epsilon_n^\uparrow)^2},\label{chernkubo}
\end{equation}
where $(J^a_{\bm k})_{mn}=\langle\Psi^{\epsilon_m^\uparrow}_{\bm k}|\partial_{\bm k}^a\hat{H}^{\uparrow}_{\bm k}|\Psi^{\epsilon_n^\uparrow}_{\bm k}\rangle$ is the expectation value of the current operator in the magnetic Bloch state $\Psi^{\epsilon_m^\uparrow}_{\bm k}$ of momentum $\bm k$ and energy $\epsilon_m^\uparrow$. As usual, $\epsilon_{xy}=-\epsilon_{yx}=1$ is the 2D antisymmetric tensor. The total Hall conductivity is the sum of the two spin contributions. The crucial feature is that $\sigma^{\uparrow}_{xy}(\mu)$ changes sign as the chemical potential $\mu$ is tuned, and the value of the chemical potential at which the sign change occurs depends on the strength of the displacement field. The calculation is estimated at $3^{\circ}$ twist angle at experimental feasible $10T$ fields where $\Phi_B=\pi/6$, without including the Zeeman shift \cite{Jiawei_HartreeFock}.

The results above are obtained for one spin direction. The Hofstadter butterfly is independent of the sign of the magnetic field and therefore is the same for each spin, but the Zeeman coupling shifts the spectra for spin up relative to those for spin down by $g\mu_BH$ with $g\approx 9\sim13$, so $\sigma_{xy}(\mu)=\sigma^{\uparrow}_{xy}(\mu+\frac{1}{2}g\mu_BH)+\sigma^{\downarrow}_{xy}(\mu-\frac{1}{2}g\mu_BH)$. For experimentally feasible fields $\sim 10T$ the Zeeman splitting is about $7meV$, comparable with the $3^{\circ}$ bandwidth \footnote{The bandwidth ($9t$) is estimated to be $30meV$ at $3^{\circ}$ twist angle \cite{Wang:2020aa}. In other words, the Zeeman splitting ($g\mu_BH\sim7meV$) is about $2t$.}. The tunable Hall sign change is plotted in Fig.~\ref{fig:ubutterfly} (f), at fixed external magnetic field and experimentally accessible chemical potential near the top of the valence band with Zeeman energy included. Further reducing the twist angle helps to observe the Hall sign reversal phenomenon, as it increases the critical chemical potential at which the reversal occurs.

\begin{figure*}[ht]
\centering
\includegraphics[width=0.75\linewidth]{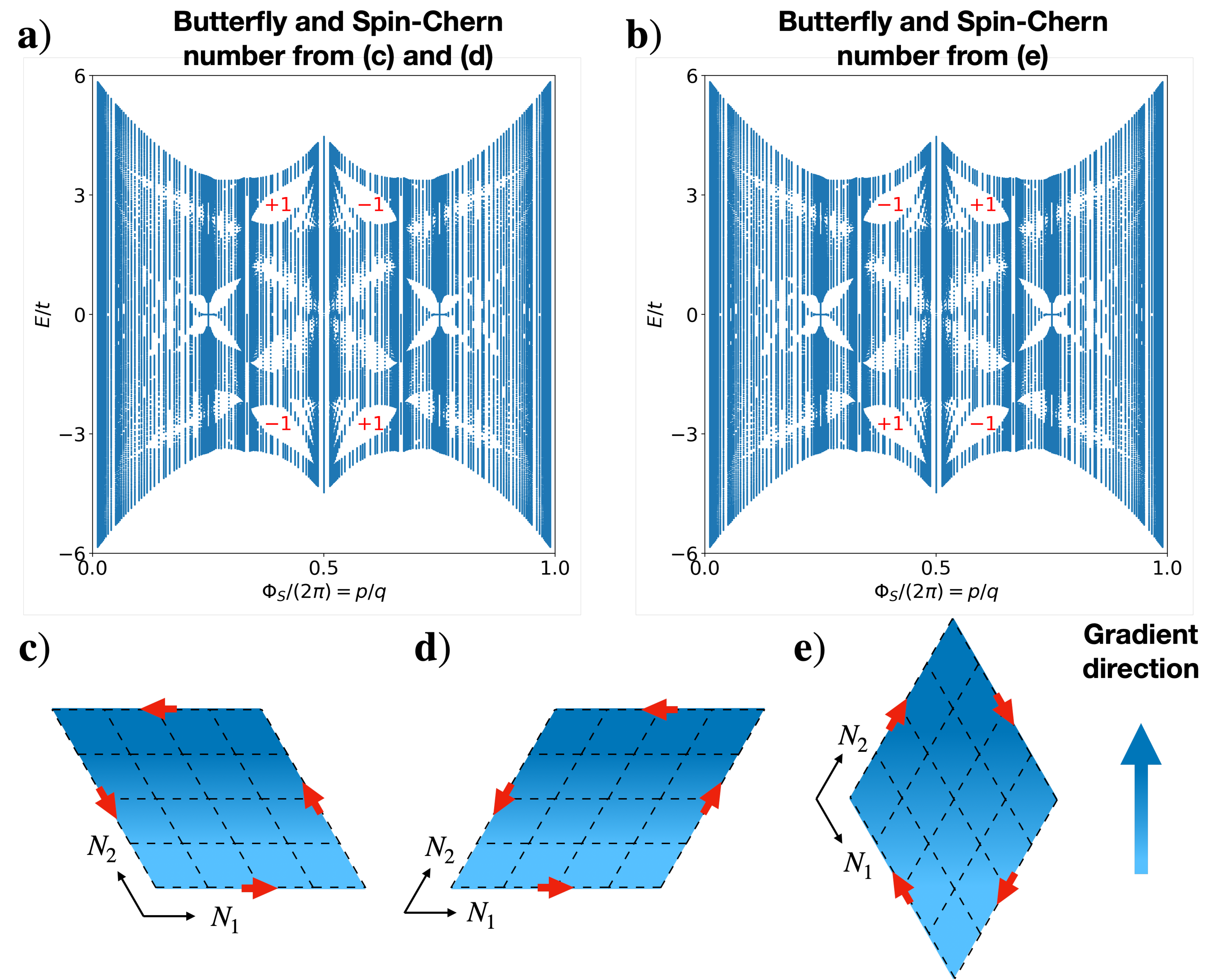}
\caption{(a) and (b): The spin-up branch of the energy spectra in the presence of a uniform gradient of displacement field in the vertical direction. The energy spectrum of (a) and (b) are plotted in terms of the displacement field gradient $\phi$, which resembles similarities to the famous Hofstadter butterfly. The spectrum is calculated with closed-boundary conditions on a torus with $(N_1,N_2)$ unit cells in each of the primitive directions. The spectrum is insensitive to the initial values $\phi_{i}(0)$ and the choice of unit cell, and exhibit an emergent particle-hole symmetry. Spin Chern numbers $C_{\uparrow}$ are marked in the four large central gaps and depend on the geometry of the torus: (a) is calculated in the geometry shown in (c) or (d); and (b) is calculated in the geometry shown in (e). The spin Chern number implies a chiral spin edge current whose direction also depends on the geometry and is shown by the red arrows in (c), (d), (e).}\label{fig:butterfly}
\end{figure*}

\textbf{Displacement field gradient induced spin current.}
We next consider the effect of a spatially inhomogeneous displacement field in the absence of an external magnetic field. The key observation is that in the presence of a spatially dependent displacement field the pseudo magnetic field in two adjacent triangular plaquettes no longer averages to zero. As an example, we consider a displacement field with a constant spatial gradient, which, without loss of generality, we assume is perpendicular to the lattice vector $\bm a_1$. For our numerical studies we have considered the three geometries shown in panels (c), (d), (e) in Fig.~\ref{fig:butterfly} with the gradient of displacement field directed vertically as shown. We first consider the non-interacting limit with $U=0$. The Hamiltonian is given by Eqn.~(\ref{moirehubbard2}) but the phase fields now have a spatial coordinate dependence:
\begin{equation}
\phi_{\bm r,\bm r+\bm a_n} = \phi_{n} + \Phi_S\cdot y,\quad n=1,2,3,\label{simplegradient}
\end{equation}
where $\bm r=x\bm a_1+y\bm a_2$, $\phi_n$ are the initial values at $y=0$.

A simple counting shows that the modulus of the net pseudo magnetic field is $|\Phi_S|$ in any primitive unit cell (any two adjacent triangles). However, its sign depends on the orientation of the unit cell: the pseudo magnetic field for spin up is negative for the geometry shown in Fig.~\ref{fig:butterfly} (c), (d) and positive for the geometry of Fig.~\ref{fig:butterfly} (e).

The Hamiltonian with the phase specified in Eqn.~(\ref{simplegradient}) is easily diagonalized with periodic boundary conditions in an enlarged unit cell containing $q$ primitive unit cells such that the phase changes by an integer multiple of $2\pi$ across the cell. The energy spectrum of the spin-up branch is plotted in Fig.~\ref{fig:butterfly} as a function of the gradient $\Phi_S=2\pi p/q$. These spectra display similarities to the standard Hofstadter butterfly, but are distinguished in a couple of aspects: first, in contrast to the magnetic field induced butterfly, in the gradient induced butterfly Landau fans occur at large pseudo magnetic field when $\Phi_S \approx \pi$ rather than small field values; second, the butterflies in Fig.~\ref{fig:butterfly} have an emergent particle-hole symmetry, and are insensitive to the initial values $\phi_{n=1,2,3}$.

The nontrivial topology induced by the gradient is determined by the Chern numbers of the butterfly spectrum, {\it i.e.} the spin Chern numbers, which are marked for the largest four gaps near the middle of the spectrum and computed following the Kubo formula in Eqn.~(\ref{chernkubo}). Since time-reversal symmetry is preserved by the displacement field, the spin down branch has opposite Chern number. As a result, when the chemical potential and $\phi$ are within these gaps, we expect chiral spin up current to travel in one direction around the sample edge, with opposite current for  the spin down edge mode. Therefore a nonzero spin current is expected. However, this spin current is not quantized and not necessarily protected because a jagged edge changes the balance between triangles with opposite values of staggered fluxes. More detailed discussion of jagged geometries is left to a future paper.

\begin{figure*}[ht]
\centering
\includegraphics[width=1.0\linewidth]{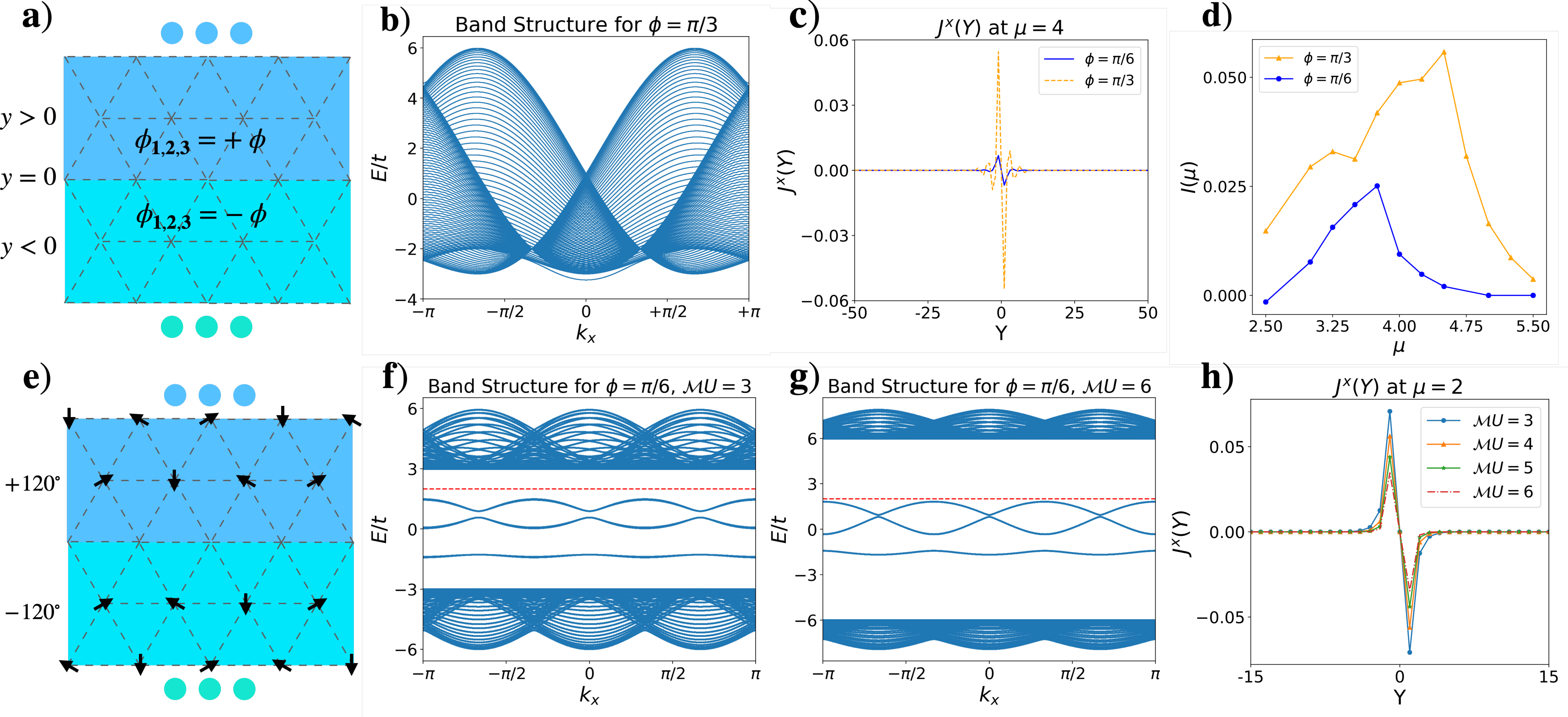}
\caption{Band structure and interface current in the junction geometry with and without magnetic order. (a) to (d) considers the non-interacting case without magnetic order. (a) The geometry of the heterostructure, where the displacement fields are uniform in the upper and lower sides of the junction. (b) The band structure ($\phi=\pi/3$) plotted in terms of the horizontal momentum $k_x$, which consists of two fans of continuum states and one isolated state near the band bottom. (c) The interface current calculated when the chemical potential $\mu/t=4$ for two different values of $\phi$. The integrated current $I(\mu)$, defined as total current over the entire lower side, as a function of chemical potential is plotted in (d) for these two values of $\phi$. The magnetic order is assumed nonzero from (e) to (h). The Hartree-Fock bands are shown in (f) and (g), where the red dotted line is the chemical potential above which the current density are plotted in (h). The current is measured in units of $te/\hbar$.}\label{fig:junc}
\end{figure*}

\textbf{Spin current at the interface.}
We now consider an abrupt change in the displacement field. We focus on the junction geometry shown in Fig.~\ref{fig:junc} (a), consisting of a sample that is infinite in the vertical direction, periodic in the horizontal direction, and characterized by a displacement field that abruptly changes sign. We parametrize this displacement field by:
\begin{eqnarray}
    \text{At $y\neq0$:}&&\quad\phi_{\bm r,\bm r+\bm a_n} = \sign(y)\phi,\quad n=1,2,3,\label{phase_interface1}\\
    \text{At $y=0$:}&&\quad\phi_{1} = 0,\quad\phi_{2}=\phi,\quad\phi_{3} = -\phi,\label{phase_interface2}
\end{eqnarray}
where $\sign(y>0)=+1$ and $\sign(y<0)=-1$ is the sign function.

The band structure is shown in Fig.~\ref{fig:junc} (b), which typically consists of two fans of continuum states connected by the interface isolated states. We also computed the current density, contributed from all Bloch states below the chemical potential $\mu$,
\begin{equation}
    J^x(\mu,y) = \int_{-\pi}^{\pi}\frac{dk_x}{2\pi}\int^{\mu}_{-\infty} d\omega^{\uparrow}\langle\Psi^{\omega^{\uparrow}}_{k_x}(y)|\hat{J}^x_{k_x}|\Psi^{\omega^{\uparrow}}_{k_x}(y)\rangle,
\end{equation}
where $\hat{J}^x_{k_x}\equiv\partial_{k_x}\hat{H}^{\uparrow}_{k_x}$ is the current operator and $\Psi^{\omega^{\uparrow}}_{k_x}$ is the Bloch state of momentum $k_x$ and energy $\omega^{\uparrow}$. The computed current exhibits a peak at the interface, as shown in Fig.~\ref{fig:junc} (c). The spin down branch contributes an opposite amount of current due to the time-reversal symmetry. A net result is zero electronic current, but sharp spin current at the interface. The spin current contributed from hole carriers near the band top is particularly important to experiments. The integrated current, defined as the current summed over the entire lower half of the junction $I(\mu)=\int_{-\infty}^0dy~J^{x}(\mu,y)$ \footnote{Current integrated over the upper junction is just opposite, because the net current of the entire sample is zero.}, is plotted in Fig.~\ref{fig:junc} (d) as a function of the chemical potential $\mu$ near the band top. Note that as in the previous case, the spin current is neither quantized nor protected from back scattering in this metallic phase, as it could be destroyed by impurities that couple states of same energy but different $k_x$.

Recent calculations show that at certain values of the displacement field the ground state of the half filled model possesses $\pm120^{\circ}$ spiral  anti-ferromagnetic order with the sign of the spiral depending on the sign of the displacement field  \cite{Jiawei_HartreeFock}. Magnetic order is most strongly favored  when $\phi=\pm\pi/6$ due to the perfect nesting of the Fermi surface. Motivated by this, we consider the presence of magnetic order, which assume is $+120^{\circ}$ order on the upper plane where $\phi=+\pi/6$, and $-120^{\circ}$ order on the lower plane where $\phi=-\pi/6$. At the interface, the magnetic order is set zero. See Fig.~\ref{fig:junc} (e) for the configuration. The magnetic order opens an energy gap in the Hartree-Fock band within which are interface localized modes, shown in (f) and (g) of Fig.~\ref{fig:junc}. Similarly, the spin currents are numerically computed in FIG.~\ref{fig:junc} (h). Thus a sharp junction between two insulating states will result in spin current protected by the magnetic gap against back-scattering.

\textbf{Conclusion.}
The low energy physics of electrons in twisted \wsetwo involves a staggered flux, of magnitude easily tunable experimentally via variation of the potential difference between the layers. We have shown that this flux leads to remarkable experimental consequences, including tunable Hofstadter butterfly spectra, tunable sign reversals of the Hall coefficient, and spin currents at sample edges and interface. The spin currents may be useful for new classes of spintronic devices. More generally, since internal magnetic fields are generic to twisted \mr systems, our work motivates further investigations of observable consequences of these fields.

\begin{acknowledgments}
J.C., J.Z. and A.J.M acknowledge support from the NSF MRSEC program through the Center for Precision-Assembled Quantum Materials (PAQM) - DMR-2011738. The Flatiron Institute is a division of the Simons Foundation.
\end{acknowledgments}

\bibliography{tTMD.bib}

\begin{thebibliography}{36}%
\makeatletter
\providecommand \@ifxundefined [1]{%
 \@ifx{#1\undefined}
}%
\providecommand \@ifnum [1]{%
 \ifnum #1\expandafter \@firstoftwo
 \else \expandafter \@secondoftwo
 \fi
}%
\providecommand \@ifx [1]{%
 \ifx #1\expandafter \@firstoftwo
 \else \expandafter \@secondoftwo
 \fi
}%
\providecommand \natexlab [1]{#1}%
\providecommand \enquote  [1]{``#1''}%
\providecommand \bibnamefont  [1]{#1}%
\providecommand \bibfnamefont [1]{#1}%
\providecommand \citenamefont [1]{#1}%
\providecommand \href@noop [0]{\@secondoftwo}%
\providecommand \href [0]{\begingroup \@sanitize@url \@href}%
\providecommand \@href[1]{\@@startlink{#1}\@@href}%
\providecommand \@@href[1]{\endgroup#1\@@endlink}%
\providecommand \@sanitize@url [0]{\catcode `\\12\catcode `\$12\catcode
  `\&12\catcode `\#12\catcode `\^12\catcode `\_12\catcode `\%12\relax}%
\providecommand \@@startlink[1]{}%
\providecommand \@@endlink[0]{}%
\providecommand \url  [0]{\begingroup\@sanitize@url \@url }%
\providecommand \@url [1]{\endgroup\@href {#1}{\urlprefix }}%
\providecommand \urlprefix  [0]{URL }%
\providecommand \Eprint [0]{\href }%
\providecommand \doibase [0]{http://dx.doi.org/}%
\providecommand \selectlanguage [0]{\@gobble}%
\providecommand \bibinfo  [0]{\@secondoftwo}%
\providecommand \bibfield  [0]{\@secondoftwo}%
\providecommand \translation [1]{[#1]}%
\providecommand \BibitemOpen [0]{}%
\providecommand \bibitemStop [0]{}%
\providecommand \bibitemNoStop [0]{.\EOS\space}%
\providecommand \EOS [0]{\spacefactor3000\relax}%
\providecommand \BibitemShut  [1]{\csname bibitem#1\endcsname}%
\let\auto@bib@innerbib\@empty
%</preamble>
\bibitem [{\citenamefont {Bistritzer}\ and\ \citenamefont
  {MacDonald}(2011)}]{Bistritzer12233}%
  \BibitemOpen
  \bibfield  {author} {\bibinfo {author} {\bibfnamefont {R.}~\bibnamefont
  {Bistritzer}}\ and\ \bibinfo {author} {\bibfnamefont {A.~H.}\ \bibnamefont
  {MacDonald}},\ }\href {\doibase 10.1073/pnas.1108174108} {\bibfield
  {journal} {\bibinfo  {journal} {Proceedings of the National Academy of
  Sciences}\ }\textbf {\bibinfo {volume} {108}},\ \bibinfo {pages} {12233}
  (\bibinfo {year} {2011})},\ \Eprint
  {http://arxiv.org/abs/https://www.pnas.org/content/108/30/12233.full.pdf}
  {https://www.pnas.org/content/108/30/12233.full.pdf} \BibitemShut {NoStop}%
\bibitem [{\citenamefont {Cao}\ \emph {et~al.}(2018{\natexlab{a}})\citenamefont
  {Cao}, \citenamefont {Fatemi}, \citenamefont {Fang}, \citenamefont
  {Watanabe}, \citenamefont {Taniguchi}, \citenamefont {Kaxiras},\ and\
  \citenamefont {Jarillo-Herrero}}]{Cao:2018aa}%
  \BibitemOpen
  \bibfield  {author} {\bibinfo {author} {\bibfnamefont {Y.}~\bibnamefont
  {Cao}}, \bibinfo {author} {\bibfnamefont {V.}~\bibnamefont {Fatemi}},
  \bibinfo {author} {\bibfnamefont {S.}~\bibnamefont {Fang}}, \bibinfo {author}
  {\bibfnamefont {K.}~\bibnamefont {Watanabe}}, \bibinfo {author}
  {\bibfnamefont {T.}~\bibnamefont {Taniguchi}}, \bibinfo {author}
  {\bibfnamefont {E.}~\bibnamefont {Kaxiras}}, \ and\ \bibinfo {author}
  {\bibfnamefont {P.}~\bibnamefont {Jarillo-Herrero}},\ }\href {\doibase
  10.1038/nature26160} {\bibfield  {journal} {\bibinfo  {journal} {Nature}\
  }\textbf {\bibinfo {volume} {556}},\ \bibinfo {pages} {43} (\bibinfo {year}
  {2018}{\natexlab{a}})}\BibitemShut {NoStop}%
\bibitem [{\citenamefont {Cao}\ \emph {et~al.}(2018{\natexlab{b}})\citenamefont
  {Cao}, \citenamefont {Fatemi}, \citenamefont {Demir}, \citenamefont {Fang},
  \citenamefont {Tomarken}, \citenamefont {Luo}, \citenamefont
  {Sanchez-Yamagishi}, \citenamefont {Watanabe}, \citenamefont {Taniguchi},
  \citenamefont {Kaxiras}, \citenamefont {Ashoori},\ and\ \citenamefont
  {Jarillo-Herrero}}]{Cao:2018ab}%
  \BibitemOpen
  \bibfield  {author} {\bibinfo {author} {\bibfnamefont {Y.}~\bibnamefont
  {Cao}}, \bibinfo {author} {\bibfnamefont {V.}~\bibnamefont {Fatemi}},
  \bibinfo {author} {\bibfnamefont {A.}~\bibnamefont {Demir}}, \bibinfo
  {author} {\bibfnamefont {S.}~\bibnamefont {Fang}}, \bibinfo {author}
  {\bibfnamefont {S.~L.}\ \bibnamefont {Tomarken}}, \bibinfo {author}
  {\bibfnamefont {J.~Y.}\ \bibnamefont {Luo}}, \bibinfo {author} {\bibfnamefont
  {J.~D.}\ \bibnamefont {Sanchez-Yamagishi}}, \bibinfo {author} {\bibfnamefont
  {K.}~\bibnamefont {Watanabe}}, \bibinfo {author} {\bibfnamefont
  {T.}~\bibnamefont {Taniguchi}}, \bibinfo {author} {\bibfnamefont
  {E.}~\bibnamefont {Kaxiras}}, \bibinfo {author} {\bibfnamefont {R.~C.}\
  \bibnamefont {Ashoori}}, \ and\ \bibinfo {author} {\bibfnamefont
  {P.}~\bibnamefont {Jarillo-Herrero}},\ }\href {\doibase 10.1038/nature26154}
  {\bibfield  {journal} {\bibinfo  {journal} {Nature}\ }\textbf {\bibinfo
  {volume} {556}},\ \bibinfo {pages} {80} (\bibinfo {year}
  {2018}{\natexlab{b}})}\BibitemShut {NoStop}%
\bibitem [{\citenamefont {Sharpe}\ \emph {et~al.}(2019)\citenamefont {Sharpe},
  \citenamefont {Fox}, \citenamefont {Barnard}, \citenamefont {Finney},
  \citenamefont {Watanabe}, \citenamefont {Taniguchi}, \citenamefont
  {Kastner},\ and\ \citenamefont {Goldhaber-Gordon}}]{Sharpe605}%
  \BibitemOpen
  \bibfield  {author} {\bibinfo {author} {\bibfnamefont {A.~L.}\ \bibnamefont
  {Sharpe}}, \bibinfo {author} {\bibfnamefont {E.~J.}\ \bibnamefont {Fox}},
  \bibinfo {author} {\bibfnamefont {A.~W.}\ \bibnamefont {Barnard}}, \bibinfo
  {author} {\bibfnamefont {J.}~\bibnamefont {Finney}}, \bibinfo {author}
  {\bibfnamefont {K.}~\bibnamefont {Watanabe}}, \bibinfo {author}
  {\bibfnamefont {T.}~\bibnamefont {Taniguchi}}, \bibinfo {author}
  {\bibfnamefont {M.~A.}\ \bibnamefont {Kastner}}, \ and\ \bibinfo {author}
  {\bibfnamefont {D.}~\bibnamefont {Goldhaber-Gordon}},\ }\href {\doibase
  10.1126/science.aaw3780} {\bibfield  {journal} {\bibinfo  {journal}
  {Science}\ }\textbf {\bibinfo {volume} {365}},\ \bibinfo {pages} {605}
  (\bibinfo {year} {2019})}\BibitemShut {NoStop}%
\bibitem [{\citenamefont {Serlin}\ \emph {et~al.}(2020)\citenamefont {Serlin},
  \citenamefont {Tschirhart}, \citenamefont {Polshyn}, \citenamefont {Zhang},
  \citenamefont {Zhu}, \citenamefont {Watanabe}, \citenamefont {Taniguchi},
  \citenamefont {Balents},\ and\ \citenamefont {Young}}]{Serlin900}%
  \BibitemOpen
  \bibfield  {author} {\bibinfo {author} {\bibfnamefont {M.}~\bibnamefont
  {Serlin}}, \bibinfo {author} {\bibfnamefont {C.~L.}\ \bibnamefont
  {Tschirhart}}, \bibinfo {author} {\bibfnamefont {H.}~\bibnamefont {Polshyn}},
  \bibinfo {author} {\bibfnamefont {Y.}~\bibnamefont {Zhang}}, \bibinfo
  {author} {\bibfnamefont {J.}~\bibnamefont {Zhu}}, \bibinfo {author}
  {\bibfnamefont {K.}~\bibnamefont {Watanabe}}, \bibinfo {author}
  {\bibfnamefont {T.}~\bibnamefont {Taniguchi}}, \bibinfo {author}
  {\bibfnamefont {L.}~\bibnamefont {Balents}}, \ and\ \bibinfo {author}
  {\bibfnamefont {A.~F.}\ \bibnamefont {Young}},\ }\href {\doibase
  10.1126/science.aay5533} {\bibfield  {journal} {\bibinfo  {journal}
  {Science}\ }\textbf {\bibinfo {volume} {367}},\ \bibinfo {pages} {900}
  (\bibinfo {year} {2020})}\BibitemShut {NoStop}%
\bibitem [{\citenamefont {Stepanov}\ \emph {et~al.}(2020)\citenamefont
  {Stepanov}, \citenamefont {Das}, \citenamefont {Lu}, \citenamefont
  {Fahimniya}, \citenamefont {Watanabe}, \citenamefont {Taniguchi},
  \citenamefont {Koppens}, \citenamefont {Lischner}, \citenamefont {Levitov},\
  and\ \citenamefont {Efetov}}]{Stepanov:2020aa}%
  \BibitemOpen
  \bibfield  {author} {\bibinfo {author} {\bibfnamefont {P.}~\bibnamefont
  {Stepanov}}, \bibinfo {author} {\bibfnamefont {I.}~\bibnamefont {Das}},
  \bibinfo {author} {\bibfnamefont {X.}~\bibnamefont {Lu}}, \bibinfo {author}
  {\bibfnamefont {A.}~\bibnamefont {Fahimniya}}, \bibinfo {author}
  {\bibfnamefont {K.}~\bibnamefont {Watanabe}}, \bibinfo {author}
  {\bibfnamefont {T.}~\bibnamefont {Taniguchi}}, \bibinfo {author}
  {\bibfnamefont {F.~H.~L.}\ \bibnamefont {Koppens}}, \bibinfo {author}
  {\bibfnamefont {J.}~\bibnamefont {Lischner}}, \bibinfo {author}
  {\bibfnamefont {L.}~\bibnamefont {Levitov}}, \ and\ \bibinfo {author}
  {\bibfnamefont {D.~K.}\ \bibnamefont {Efetov}},\ }\href {\doibase
  10.1038/s41586-020-2459-6} {\bibfield  {journal} {\bibinfo  {journal}
  {Nature}\ }\textbf {\bibinfo {volume} {583}},\ \bibinfo {pages} {375}
  (\bibinfo {year} {2020})}\BibitemShut {NoStop}%
\bibitem [{\citenamefont {Tang}\ \emph {et~al.}(2020)\citenamefont {Tang},
  \citenamefont {Li}, \citenamefont {Li}, \citenamefont {Xu}, \citenamefont
  {Liu}, \citenamefont {Barmak}, \citenamefont {Watanabe}, \citenamefont
  {Taniguchi}, \citenamefont {MacDonald}, \citenamefont {Shan},\ and\
  \citenamefont {Mak}}]{Tang:2020aa}%
  \BibitemOpen
  \bibfield  {author} {\bibinfo {author} {\bibfnamefont {Y.}~\bibnamefont
  {Tang}}, \bibinfo {author} {\bibfnamefont {L.}~\bibnamefont {Li}}, \bibinfo
  {author} {\bibfnamefont {T.}~\bibnamefont {Li}}, \bibinfo {author}
  {\bibfnamefont {Y.}~\bibnamefont {Xu}}, \bibinfo {author} {\bibfnamefont
  {S.}~\bibnamefont {Liu}}, \bibinfo {author} {\bibfnamefont {K.}~\bibnamefont
  {Barmak}}, \bibinfo {author} {\bibfnamefont {K.}~\bibnamefont {Watanabe}},
  \bibinfo {author} {\bibfnamefont {T.}~\bibnamefont {Taniguchi}}, \bibinfo
  {author} {\bibfnamefont {A.~H.}\ \bibnamefont {MacDonald}}, \bibinfo {author}
  {\bibfnamefont {J.}~\bibnamefont {Shan}}, \ and\ \bibinfo {author}
  {\bibfnamefont {K.~F.}\ \bibnamefont {Mak}},\ }\href {\doibase
  10.1038/s41586-020-2085-3} {\bibfield  {journal} {\bibinfo  {journal}
  {Nature}\ }\textbf {\bibinfo {volume} {579}},\ \bibinfo {pages} {353}
  (\bibinfo {year} {2020})}\BibitemShut {NoStop}%
\bibitem [{\citenamefont {Wang}\ \emph {et~al.}(2020)\citenamefont {Wang},
  \citenamefont {Shih}, \citenamefont {Ghiotto}, \citenamefont {Xian},
  \citenamefont {Rhodes}, \citenamefont {Tan}, \citenamefont {Claassen},
  \citenamefont {Kennes}, \citenamefont {Bai}, \citenamefont {Kim},
  \citenamefont {Watanabe}, \citenamefont {Taniguchi}, \citenamefont {Zhu},
  \citenamefont {Hone}, \citenamefont {Rubio}, \citenamefont {Pasupathy},\ and\
  \citenamefont {Dean}}]{Wang:2020aa}%
  \BibitemOpen
  \bibfield  {author} {\bibinfo {author} {\bibfnamefont {L.}~\bibnamefont
  {Wang}}, \bibinfo {author} {\bibfnamefont {E.-M.}\ \bibnamefont {Shih}},
  \bibinfo {author} {\bibfnamefont {A.}~\bibnamefont {Ghiotto}}, \bibinfo
  {author} {\bibfnamefont {L.}~\bibnamefont {Xian}}, \bibinfo {author}
  {\bibfnamefont {D.~A.}\ \bibnamefont {Rhodes}}, \bibinfo {author}
  {\bibfnamefont {C.}~\bibnamefont {Tan}}, \bibinfo {author} {\bibfnamefont
  {M.}~\bibnamefont {Claassen}}, \bibinfo {author} {\bibfnamefont {D.~M.}\
  \bibnamefont {Kennes}}, \bibinfo {author} {\bibfnamefont {Y.}~\bibnamefont
  {Bai}}, \bibinfo {author} {\bibfnamefont {B.}~\bibnamefont {Kim}}, \bibinfo
  {author} {\bibfnamefont {K.}~\bibnamefont {Watanabe}}, \bibinfo {author}
  {\bibfnamefont {T.}~\bibnamefont {Taniguchi}}, \bibinfo {author}
  {\bibfnamefont {X.}~\bibnamefont {Zhu}}, \bibinfo {author} {\bibfnamefont
  {J.}~\bibnamefont {Hone}}, \bibinfo {author} {\bibfnamefont {A.}~\bibnamefont
  {Rubio}}, \bibinfo {author} {\bibfnamefont {A.~N.}\ \bibnamefont
  {Pasupathy}}, \ and\ \bibinfo {author} {\bibfnamefont {C.~R.}\ \bibnamefont
  {Dean}},\ }\href {\doibase 10.1038/s41563-020-0708-6} {\bibfield  {journal}
  {\bibinfo  {journal} {Nature Materials}\ }\textbf {\bibinfo {volume} {19}},\
  \bibinfo {pages} {861} (\bibinfo {year} {2020})}\BibitemShut {NoStop}%
\bibitem [{\citenamefont {Li}\ \emph {et~al.}(2021)\citenamefont {Li},
  \citenamefont {Jiang}, \citenamefont {Li}, \citenamefont {Zhang},
  \citenamefont {Kang}, \citenamefont {Zhu}, \citenamefont {Watanabe},
  \citenamefont {Taniguchi}, \citenamefont {Chowdhury}, \citenamefont {Fu},
  \citenamefont {Shan},\ and\ \citenamefont {Mak}}]{Li:2021aa}%
  \BibitemOpen
  \bibfield  {author} {\bibinfo {author} {\bibfnamefont {T.}~\bibnamefont
  {Li}}, \bibinfo {author} {\bibfnamefont {S.}~\bibnamefont {Jiang}}, \bibinfo
  {author} {\bibfnamefont {L.}~\bibnamefont {Li}}, \bibinfo {author}
  {\bibfnamefont {Y.}~\bibnamefont {Zhang}}, \bibinfo {author} {\bibfnamefont
  {K.}~\bibnamefont {Kang}}, \bibinfo {author} {\bibfnamefont {J.}~\bibnamefont
  {Zhu}}, \bibinfo {author} {\bibfnamefont {K.}~\bibnamefont {Watanabe}},
  \bibinfo {author} {\bibfnamefont {T.}~\bibnamefont {Taniguchi}}, \bibinfo
  {author} {\bibfnamefont {D.}~\bibnamefont {Chowdhury}}, \bibinfo {author}
  {\bibfnamefont {L.}~\bibnamefont {Fu}}, \bibinfo {author} {\bibfnamefont
  {J.}~\bibnamefont {Shan}}, \ and\ \bibinfo {author} {\bibfnamefont {K.~F.}\
  \bibnamefont {Mak}},\ }\href {\doibase 10.1038/s41586-021-03853-0} {\bibfield
   {journal} {\bibinfo  {journal} {Nature}\ }\textbf {\bibinfo {volume}
  {597}},\ \bibinfo {pages} {350} (\bibinfo {year} {2021})}\BibitemShut
  {NoStop}%
\bibitem [{\citenamefont {Ghiotto}\ \emph {et~al.}(2021)\citenamefont
  {Ghiotto}, \citenamefont {Shih}, \citenamefont {Pereira}, \citenamefont
  {Rhodes}, \citenamefont {Kim}, \citenamefont {Zang}, \citenamefont {Millis},
  \citenamefont {Watanabe}, \citenamefont {Taniguchi}, \citenamefont {Hone},
  \citenamefont {Wang}, \citenamefont {Dean},\ and\ \citenamefont
  {Pasupathy}}]{Ghiotto:2021aa}%
  \BibitemOpen
  \bibfield  {author} {\bibinfo {author} {\bibfnamefont {A.}~\bibnamefont
  {Ghiotto}}, \bibinfo {author} {\bibfnamefont {E.-M.}\ \bibnamefont {Shih}},
  \bibinfo {author} {\bibfnamefont {G.~S. S.~G.}\ \bibnamefont {Pereira}},
  \bibinfo {author} {\bibfnamefont {D.~A.}\ \bibnamefont {Rhodes}}, \bibinfo
  {author} {\bibfnamefont {B.}~\bibnamefont {Kim}}, \bibinfo {author}
  {\bibfnamefont {J.}~\bibnamefont {Zang}}, \bibinfo {author} {\bibfnamefont
  {A.~J.}\ \bibnamefont {Millis}}, \bibinfo {author} {\bibfnamefont
  {K.}~\bibnamefont {Watanabe}}, \bibinfo {author} {\bibfnamefont
  {T.}~\bibnamefont {Taniguchi}}, \bibinfo {author} {\bibfnamefont {J.~C.}\
  \bibnamefont {Hone}}, \bibinfo {author} {\bibfnamefont {L.}~\bibnamefont
  {Wang}}, \bibinfo {author} {\bibfnamefont {C.~R.}\ \bibnamefont {Dean}}, \
  and\ \bibinfo {author} {\bibfnamefont {A.~N.}\ \bibnamefont {Pasupathy}},\
  }\href {\doibase 10.1038/s41586-021-03815-6} {\bibfield  {journal} {\bibinfo
  {journal} {Nature}\ }\textbf {\bibinfo {volume} {597}},\ \bibinfo {pages}
  {345} (\bibinfo {year} {2021})}\BibitemShut {NoStop}%
\bibitem [{\citenamefont {Jin}\ \emph {et~al.}(2021)\citenamefont {Jin},
  \citenamefont {Tao}, \citenamefont {Li}, \citenamefont {Xu}, \citenamefont
  {Tang}, \citenamefont {Zhu}, \citenamefont {Liu}, \citenamefont {Watanabe},
  \citenamefont {Taniguchi}, \citenamefont {Hone}, \citenamefont {Fu},
  \citenamefont {Shan},\ and\ \citenamefont {Mak}}]{Jin:2021aa}%
  \BibitemOpen
  \bibfield  {author} {\bibinfo {author} {\bibfnamefont {C.}~\bibnamefont
  {Jin}}, \bibinfo {author} {\bibfnamefont {Z.}~\bibnamefont {Tao}}, \bibinfo
  {author} {\bibfnamefont {T.}~\bibnamefont {Li}}, \bibinfo {author}
  {\bibfnamefont {Y.}~\bibnamefont {Xu}}, \bibinfo {author} {\bibfnamefont
  {Y.}~\bibnamefont {Tang}}, \bibinfo {author} {\bibfnamefont {J.}~\bibnamefont
  {Zhu}}, \bibinfo {author} {\bibfnamefont {S.}~\bibnamefont {Liu}}, \bibinfo
  {author} {\bibfnamefont {K.}~\bibnamefont {Watanabe}}, \bibinfo {author}
  {\bibfnamefont {T.}~\bibnamefont {Taniguchi}}, \bibinfo {author}
  {\bibfnamefont {J.~C.}\ \bibnamefont {Hone}}, \bibinfo {author}
  {\bibfnamefont {L.}~\bibnamefont {Fu}}, \bibinfo {author} {\bibfnamefont
  {J.}~\bibnamefont {Shan}}, \ and\ \bibinfo {author} {\bibfnamefont {K.~F.}\
  \bibnamefont {Mak}},\ }\href {\doibase 10.1038/s41563-021-00959-8} {\bibfield
   {journal} {\bibinfo  {journal} {Nature Materials}\ }\textbf {\bibinfo
  {volume} {20}},\ \bibinfo {pages} {940} (\bibinfo {year} {2021})}\BibitemShut
  {NoStop}%
\bibitem [{\citenamefont {Andersen}\ \emph {et~al.}(2021)\citenamefont
  {Andersen}, \citenamefont {Scuri}, \citenamefont {Sushko}, \citenamefont
  {De~Greve}, \citenamefont {Sung}, \citenamefont {Zhou}, \citenamefont {Wild},
  \citenamefont {Gelly}, \citenamefont {Heo}, \citenamefont {B{\'e}rub{\'e}},
  \citenamefont {Joe}, \citenamefont {Jauregui}, \citenamefont {Watanabe},
  \citenamefont {Taniguchi}, \citenamefont {Kim}, \citenamefont {Park},\ and\
  \citenamefont {Lukin}}]{Andersen:2021aa}%
  \BibitemOpen
  \bibfield  {author} {\bibinfo {author} {\bibfnamefont {T.~I.}\ \bibnamefont
  {Andersen}}, \bibinfo {author} {\bibfnamefont {G.}~\bibnamefont {Scuri}},
  \bibinfo {author} {\bibfnamefont {A.}~\bibnamefont {Sushko}}, \bibinfo
  {author} {\bibfnamefont {K.}~\bibnamefont {De~Greve}}, \bibinfo {author}
  {\bibfnamefont {J.}~\bibnamefont {Sung}}, \bibinfo {author} {\bibfnamefont
  {Y.}~\bibnamefont {Zhou}}, \bibinfo {author} {\bibfnamefont {D.~S.}\
  \bibnamefont {Wild}}, \bibinfo {author} {\bibfnamefont {R.~J.}\ \bibnamefont
  {Gelly}}, \bibinfo {author} {\bibfnamefont {H.}~\bibnamefont {Heo}}, \bibinfo
  {author} {\bibfnamefont {D.}~\bibnamefont {B{\'e}rub{\'e}}}, \bibinfo
  {author} {\bibfnamefont {A.~Y.}\ \bibnamefont {Joe}}, \bibinfo {author}
  {\bibfnamefont {L.~A.}\ \bibnamefont {Jauregui}}, \bibinfo {author}
  {\bibfnamefont {K.}~\bibnamefont {Watanabe}}, \bibinfo {author}
  {\bibfnamefont {T.}~\bibnamefont {Taniguchi}}, \bibinfo {author}
  {\bibfnamefont {P.}~\bibnamefont {Kim}}, \bibinfo {author} {\bibfnamefont
  {H.}~\bibnamefont {Park}}, \ and\ \bibinfo {author} {\bibfnamefont {M.~D.}\
  \bibnamefont {Lukin}},\ }\href {\doibase 10.1038/s41563-020-00873-5}
  {\bibfield  {journal} {\bibinfo  {journal} {Nature Materials}\ }\textbf
  {\bibinfo {volume} {20}},\ \bibinfo {pages} {480} (\bibinfo {year}
  {2021})}\BibitemShut {NoStop}%
\bibitem [{\citenamefont {Xu}\ \emph {et~al.}(2020)\citenamefont {Xu},
  \citenamefont {Liu}, \citenamefont {Rhodes}, \citenamefont {Watanabe},
  \citenamefont {Taniguchi}, \citenamefont {Hone}, \citenamefont {Elser},
  \citenamefont {Mak},\ and\ \citenamefont {Shan}}]{Xu:2020aa}%
  \BibitemOpen
  \bibfield  {author} {\bibinfo {author} {\bibfnamefont {Y.}~\bibnamefont
  {Xu}}, \bibinfo {author} {\bibfnamefont {S.}~\bibnamefont {Liu}}, \bibinfo
  {author} {\bibfnamefont {D.~A.}\ \bibnamefont {Rhodes}}, \bibinfo {author}
  {\bibfnamefont {K.}~\bibnamefont {Watanabe}}, \bibinfo {author}
  {\bibfnamefont {T.}~\bibnamefont {Taniguchi}}, \bibinfo {author}
  {\bibfnamefont {J.}~\bibnamefont {Hone}}, \bibinfo {author} {\bibfnamefont
  {V.}~\bibnamefont {Elser}}, \bibinfo {author} {\bibfnamefont {K.~F.}\
  \bibnamefont {Mak}}, \ and\ \bibinfo {author} {\bibfnamefont
  {J.}~\bibnamefont {Shan}},\ }\href {\doibase 10.1038/s41586-020-2868-6}
  {\bibfield  {journal} {\bibinfo  {journal} {Nature}\ }\textbf {\bibinfo
  {volume} {587}},\ \bibinfo {pages} {214} (\bibinfo {year}
  {2020})}\BibitemShut {NoStop}%
\bibitem [{\citenamefont {Wu}\ \emph {et~al.}(2018)\citenamefont {Wu},
  \citenamefont {Lovorn}, \citenamefont {Tutuc},\ and\ \citenamefont
  {MacDonald}}]{FengchengWu_PRL18}%
  \BibitemOpen
  \bibfield  {author} {\bibinfo {author} {\bibfnamefont {F.}~\bibnamefont
  {Wu}}, \bibinfo {author} {\bibfnamefont {T.}~\bibnamefont {Lovorn}}, \bibinfo
  {author} {\bibfnamefont {E.}~\bibnamefont {Tutuc}}, \ and\ \bibinfo {author}
  {\bibfnamefont {A.~H.}\ \bibnamefont {MacDonald}},\ }\href {\doibase
  10.1103/PhysRevLett.121.026402} {\bibfield  {journal} {\bibinfo  {journal}
  {Phys. Rev. Lett.}\ }\textbf {\bibinfo {volume} {121}},\ \bibinfo {pages}
  {026402} (\bibinfo {year} {2018})}\BibitemShut {NoStop}%
\bibitem [{\citenamefont {Wu}\ \emph {et~al.}(2019)\citenamefont {Wu},
  \citenamefont {Lovorn}, \citenamefont {Tutuc}, \citenamefont {Martin},\ and\
  \citenamefont {MacDonald}}]{FengchengWu_PRL19}%
  \BibitemOpen
  \bibfield  {author} {\bibinfo {author} {\bibfnamefont {F.}~\bibnamefont
  {Wu}}, \bibinfo {author} {\bibfnamefont {T.}~\bibnamefont {Lovorn}}, \bibinfo
  {author} {\bibfnamefont {E.}~\bibnamefont {Tutuc}}, \bibinfo {author}
  {\bibfnamefont {I.}~\bibnamefont {Martin}}, \ and\ \bibinfo {author}
  {\bibfnamefont {A.~H.}\ \bibnamefont {MacDonald}},\ }\href {\doibase
  10.1103/PhysRevLett.122.086402} {\bibfield  {journal} {\bibinfo  {journal}
  {Phys. Rev. Lett.}\ }\textbf {\bibinfo {volume} {122}},\ \bibinfo {pages}
  {086402} (\bibinfo {year} {2019})}\BibitemShut {NoStop}%
\bibitem [{\citenamefont {Wang}\ \emph {et~al.}(2021)\citenamefont {Wang},
  \citenamefont {Yu}, \citenamefont {Kwan}, \citenamefont {Jia}, \citenamefont
  {Lei}, \citenamefont {Klemenz}, \citenamefont {Cevallos}, \citenamefont
  {Devakul}, \citenamefont {Watanabe}, \citenamefont {Taniguchi}, \citenamefont
  {Sondhi}, \citenamefont {Cava}, \citenamefont {Schoop}, \citenamefont
  {Parameswaran},\ and\ \citenamefont {Wu}}]{wang2021onedimensional}%
  \BibitemOpen
  \bibfield  {author} {\bibinfo {author} {\bibfnamefont {P.}~\bibnamefont
  {Wang}}, \bibinfo {author} {\bibfnamefont {G.}~\bibnamefont {Yu}}, \bibinfo
  {author} {\bibfnamefont {Y.~H.}\ \bibnamefont {Kwan}}, \bibinfo {author}
  {\bibfnamefont {Y.}~\bibnamefont {Jia}}, \bibinfo {author} {\bibfnamefont
  {S.}~\bibnamefont {Lei}}, \bibinfo {author} {\bibfnamefont {S.}~\bibnamefont
  {Klemenz}}, \bibinfo {author} {\bibfnamefont {F.~A.}\ \bibnamefont
  {Cevallos}}, \bibinfo {author} {\bibfnamefont {T.}~\bibnamefont {Devakul}},
  \bibinfo {author} {\bibfnamefont {K.}~\bibnamefont {Watanabe}}, \bibinfo
  {author} {\bibfnamefont {T.}~\bibnamefont {Taniguchi}}, \bibinfo {author}
  {\bibfnamefont {S.~L.}\ \bibnamefont {Sondhi}}, \bibinfo {author}
  {\bibfnamefont {R.~J.}\ \bibnamefont {Cava}}, \bibinfo {author}
  {\bibfnamefont {L.~M.}\ \bibnamefont {Schoop}}, \bibinfo {author}
  {\bibfnamefont {S.~A.}\ \bibnamefont {Parameswaran}}, \ and\ \bibinfo
  {author} {\bibfnamefont {S.}~\bibnamefont {Wu}},\ }\href@noop {} {\
  (\bibinfo {year} {2021})},\ \Eprint {http://arxiv.org/abs/2109.04637}
  {arXiv:2109.04637 [cond-mat.mes-hall]} \BibitemShut {NoStop}%
\bibitem [{\citenamefont {{Devakul}}\ \emph {et~al.}(2021)\citenamefont
  {{Devakul}}, \citenamefont {{Cr{\'e}pel}}, \citenamefont {{Zhang}},\ and\
  \citenamefont {{Fu}}}]{Liang_moireTMD21}%
  \BibitemOpen
  \bibfield  {author} {\bibinfo {author} {\bibfnamefont {T.}~\bibnamefont
  {{Devakul}}}, \bibinfo {author} {\bibfnamefont {V.}~\bibnamefont
  {{Cr{\'e}pel}}}, \bibinfo {author} {\bibfnamefont {Y.}~\bibnamefont
  {{Zhang}}}, \ and\ \bibinfo {author} {\bibfnamefont {L.}~\bibnamefont
  {{Fu}}},\ }\href@noop {} {\bibfield  {journal} {\bibinfo  {journal} {arXiv
  e-prints}\ ,\ \bibinfo {eid} {arXiv:2106.11954}} (\bibinfo {year} {2021})},\
  \Eprint {http://arxiv.org/abs/2106.11954} {arXiv:2106.11954
  [cond-mat.mes-hall]} \BibitemShut {NoStop}%
\bibitem [{\citenamefont {Zhang}\ \emph {et~al.}(2021)\citenamefont {Zhang},
  \citenamefont {Devakul},\ and\ \citenamefont {Fu}}]{Liang_spintexture21}%
  \BibitemOpen
  \bibfield  {author} {\bibinfo {author} {\bibfnamefont {Y.}~\bibnamefont
  {Zhang}}, \bibinfo {author} {\bibfnamefont {T.}~\bibnamefont {Devakul}}, \
  and\ \bibinfo {author} {\bibfnamefont {L.}~\bibnamefont {Fu}},\ }\href
  {\doibase 10.1073/pnas.2112673118} {\bibfield  {journal} {\bibinfo  {journal}
  {Proceedings of the National Academy of Sciences}\ }\textbf {\bibinfo
  {volume} {118}} (\bibinfo {year} {2021}),\ 10.1073/pnas.2112673118},\ \Eprint
  {http://arxiv.org/abs/https://www.pnas.org/content/118/36/e2112673118.full.pdf}
  {https://www.pnas.org/content/118/36/e2112673118.full.pdf} \BibitemShut
  {NoStop}%
\bibitem [{\citenamefont {{Chao Hu}}\ and\ \citenamefont
  {{MacDonald}}(2021)}]{Macdonald_TMD21}%
  \BibitemOpen
  \bibfield  {author} {\bibinfo {author} {\bibfnamefont {N.}~\bibnamefont
  {{Chao Hu}}}\ and\ \bibinfo {author} {\bibfnamefont {A.~H.}\ \bibnamefont
  {{MacDonald}}},\ }\href@noop {} {\bibfield  {journal} {\bibinfo  {journal}
  {arXiv e-prints}\ ,\ \bibinfo {eid} {arXiv:2108.02159}} (\bibinfo {year}
  {2021})},\ \Eprint {http://arxiv.org/abs/2108.02159} {arXiv:2108.02159
  [cond-mat.str-el]} \BibitemShut {NoStop}%
\bibitem [{\citenamefont {{Morales-Dur{\'a}n}}\ \emph
  {et~al.}(2021)\citenamefont {{Morales-Dur{\'a}n}}, \citenamefont {{Chao Hu}},
  \citenamefont {{Potasz}},\ and\ \citenamefont
  {{MacDonald}}}]{Macdonald_moireHubbard21}%
  \BibitemOpen
  \bibfield  {author} {\bibinfo {author} {\bibfnamefont {N.}~\bibnamefont
  {{Morales-Dur{\'a}n}}}, \bibinfo {author} {\bibfnamefont {N.}~\bibnamefont
  {{Chao Hu}}}, \bibinfo {author} {\bibfnamefont {P.}~\bibnamefont {{Potasz}}},
  \ and\ \bibinfo {author} {\bibfnamefont {A.~H.}\ \bibnamefont
  {{MacDonald}}},\ }\href@noop {} {\bibfield  {journal} {\bibinfo  {journal}
  {arXiv e-prints}\ ,\ \bibinfo {eid} {arXiv:2108.03313}} (\bibinfo {year}
  {2021})},\ \Eprint {http://arxiv.org/abs/2108.03313} {arXiv:2108.03313
  [cond-mat.str-el]} \BibitemShut {NoStop}%
\bibitem [{\citenamefont {{Schrade}}\ and\ \citenamefont
  {{Fu}}(2021)}]{Liang_SC_tTMD21}%
  \BibitemOpen
  \bibfield  {author} {\bibinfo {author} {\bibfnamefont {C.}~\bibnamefont
  {{Schrade}}}\ and\ \bibinfo {author} {\bibfnamefont {L.}~\bibnamefont
  {{Fu}}},\ }\href@noop {} {\bibfield  {journal} {\bibinfo  {journal} {arXiv
  e-prints}\ ,\ \bibinfo {eid} {arXiv:2110.10172}} (\bibinfo {year} {2021})},\
  \Eprint {http://arxiv.org/abs/2110.10172} {arXiv:2110.10172
  [cond-mat.supr-con]} \BibitemShut {NoStop}%
\bibitem [{\citenamefont {{Pan}}\ and\ \citenamefont {{Das
  Sarma}}(2021)}]{DasSarma_Int_Range21}%
  \BibitemOpen
  \bibfield  {author} {\bibinfo {author} {\bibfnamefont {H.}~\bibnamefont
  {{Pan}}}\ and\ \bibinfo {author} {\bibfnamefont {S.}~\bibnamefont {{Das
  Sarma}}},\ }\href@noop {} {\bibfield  {journal} {\bibinfo  {journal} {arXiv
  e-prints}\ ,\ \bibinfo {eid} {arXiv:2110.11330}} (\bibinfo {year} {2021})},\
  \Eprint {http://arxiv.org/abs/2110.11330} {arXiv:2110.11330
  [cond-mat.str-el]} \BibitemShut {NoStop}%
\bibitem [{\citenamefont {{Zhang}}\ and\ \citenamefont
  {{Vishwanath}}(2020)}]{yahui_electricaldetectionsplinliquid}%
  \BibitemOpen
  \bibfield  {author} {\bibinfo {author} {\bibfnamefont {Y.-H.}\ \bibnamefont
  {{Zhang}}}\ and\ \bibinfo {author} {\bibfnamefont {A.}~\bibnamefont
  {{Vishwanath}}},\ }\href@noop {} {\bibfield  {journal} {\bibinfo  {journal}
  {arXiv e-prints}\ ,\ \bibinfo {eid} {arXiv:2005.12925}} (\bibinfo {year}
  {2020})},\ \Eprint {http://arxiv.org/abs/2005.12925} {arXiv:2005.12925
  [cond-mat.str-el]} \BibitemShut {NoStop}%
\bibitem [{\citenamefont {{Kiese}}\ \emph {et~al.}(2021)\citenamefont
  {{Kiese}}, \citenamefont {{He}}, \citenamefont {{Hickey}}, \citenamefont
  {{Rubio}},\ and\ \citenamefont {{Kennes}}}]{Kennes_tTMD_spinliquid}%
  \BibitemOpen
  \bibfield  {author} {\bibinfo {author} {\bibfnamefont {D.}~\bibnamefont
  {{Kiese}}}, \bibinfo {author} {\bibfnamefont {Y.}~\bibnamefont {{He}}},
  \bibinfo {author} {\bibfnamefont {C.}~\bibnamefont {{Hickey}}}, \bibinfo
  {author} {\bibfnamefont {A.}~\bibnamefont {{Rubio}}}, \ and\ \bibinfo
  {author} {\bibfnamefont {D.~M.}\ \bibnamefont {{Kennes}}},\ }\href@noop {}
  {\bibfield  {journal} {\bibinfo  {journal} {arXiv e-prints}\ ,\ \bibinfo
  {eid} {arXiv:2110.10179}} (\bibinfo {year} {2021})},\ \Eprint
  {http://arxiv.org/abs/2110.10179} {arXiv:2110.10179 [cond-mat.str-el]}
  \BibitemShut {NoStop}%
\bibitem [{\citenamefont {Pan}\ \emph {et~al.}(2020)\citenamefont {Pan},
  \citenamefont {Wu},\ and\ \citenamefont {Das~Sarma}}]{DasSarma_tTMD_PRR}%
  \BibitemOpen
  \bibfield  {author} {\bibinfo {author} {\bibfnamefont {H.}~\bibnamefont
  {Pan}}, \bibinfo {author} {\bibfnamefont {F.}~\bibnamefont {Wu}}, \ and\
  \bibinfo {author} {\bibfnamefont {S.}~\bibnamefont {Das~Sarma}},\ }\href
  {\doibase 10.1103/PhysRevResearch.2.033087} {\bibfield  {journal} {\bibinfo
  {journal} {Phys. Rev. Research}\ }\textbf {\bibinfo {volume} {2}},\ \bibinfo
  {pages} {033087} (\bibinfo {year} {2020})}\BibitemShut {NoStop}%
\bibitem [{\citenamefont {Zang}\ \emph {et~al.}(2021)\citenamefont {Zang},
  \citenamefont {Wang}, \citenamefont {Cano},\ and\ \citenamefont
  {Millis}}]{Jiawei_HartreeFock}%
  \BibitemOpen
  \bibfield  {author} {\bibinfo {author} {\bibfnamefont {J.}~\bibnamefont
  {Zang}}, \bibinfo {author} {\bibfnamefont {J.}~\bibnamefont {Wang}}, \bibinfo
  {author} {\bibfnamefont {J.}~\bibnamefont {Cano}}, \ and\ \bibinfo {author}
  {\bibfnamefont {A.~J.}\ \bibnamefont {Millis}},\ }\href {\doibase
  10.1103/PhysRevB.104.075150} {\bibfield  {journal} {\bibinfo  {journal}
  {Phys. Rev. B}\ }\textbf {\bibinfo {volume} {104}},\ \bibinfo {pages}
  {075150} (\bibinfo {year} {2021})}\BibitemShut {NoStop}%
\bibitem [{\citenamefont {Bi}\ and\ \citenamefont {Fu}(2021)}]{Bi:2021aa}%
  \BibitemOpen
  \bibfield  {author} {\bibinfo {author} {\bibfnamefont {Z.}~\bibnamefont
  {Bi}}\ and\ \bibinfo {author} {\bibfnamefont {L.}~\bibnamefont {Fu}},\ }\href
  {\doibase 10.1038/s41467-020-20802-z} {\bibfield  {journal} {\bibinfo
  {journal} {Nature Communications}\ }\textbf {\bibinfo {volume} {12}},\
  \bibinfo {pages} {642} (\bibinfo {year} {2021})}\BibitemShut {NoStop}%
\bibitem [{\citenamefont {Manzeli}\ \emph {et~al.}(2017)\citenamefont
  {Manzeli}, \citenamefont {Ovchinnikov}, \citenamefont {Pasquier},
  \citenamefont {Yazyev},\ and\ \citenamefont {Kis}}]{Manzeli:2017aa}%
  \BibitemOpen
  \bibfield  {author} {\bibinfo {author} {\bibfnamefont {S.}~\bibnamefont
  {Manzeli}}, \bibinfo {author} {\bibfnamefont {D.}~\bibnamefont
  {Ovchinnikov}}, \bibinfo {author} {\bibfnamefont {D.}~\bibnamefont
  {Pasquier}}, \bibinfo {author} {\bibfnamefont {O.~V.}\ \bibnamefont
  {Yazyev}}, \ and\ \bibinfo {author} {\bibfnamefont {A.}~\bibnamefont {Kis}},\
  }\href {\doibase 10.1038/natrevmats.2017.33} {\bibfield  {journal} {\bibinfo
  {journal} {Nature Reviews Materials}\ }\textbf {\bibinfo {volume} {2}},\
  \bibinfo {pages} {17033} (\bibinfo {year} {2017})}\BibitemShut {NoStop}%
\bibitem [{Note1()}]{Note1}%
  \BibitemOpen
  \bibinfo {note} {The displacement field also affects the magnitude of the
  hopping but this effect is not directly relevant for our calculations and is
  not notated here.}\BibitemShut {Stop}%
\bibitem [{\citenamefont {{Hofstadter}}(1976)}]{Hofstadter_butterfly}%
  \BibitemOpen
  \bibfield  {author} {\bibinfo {author} {\bibfnamefont {D.~R.}\ \bibnamefont
  {{Hofstadter}}},\ }\href {\doibase 10.1103/PhysRevB.14.2239} {\bibfield
  {journal} {\bibinfo  {journal} {\prb}\ }\textbf {\bibinfo {volume} {14}},\
  \bibinfo {pages} {2239} (\bibinfo {year} {1976})}\BibitemShut {NoStop}%
\bibitem [{\citenamefont {Li}\ \emph {et~al.}(2011)\citenamefont {Li},
  \citenamefont {Wang},\ and\ \citenamefont {Gong}}]{Li_2011}%
  \BibitemOpen
  \bibfield  {author} {\bibinfo {author} {\bibfnamefont {J.}~\bibnamefont
  {Li}}, \bibinfo {author} {\bibfnamefont {Y.-F.}\ \bibnamefont {Wang}}, \ and\
  \bibinfo {author} {\bibfnamefont {C.-D.}\ \bibnamefont {Gong}},\ }\href
  {\doibase 10.1088/0953-8984/23/15/156002} {\bibfield  {journal} {\bibinfo
  {journal} {Journal of Physics: Condensed Matter}\ }\textbf {\bibinfo {volume}
  {23}},\ \bibinfo {pages} {156002} (\bibinfo {year} {2011})}\BibitemShut
  {NoStop}%
\bibitem [{\citenamefont {Kubo}(1957)}]{kubo1}%
  \BibitemOpen
  \bibfield  {author} {\bibinfo {author} {\bibfnamefont {R.}~\bibnamefont
  {Kubo}},\ }\href {\doibase 10.1143/JPSJ.12.570} {\bibfield  {journal}
  {\bibinfo  {journal} {Journal of the Physical Society of Japan}\ }\textbf
  {\bibinfo {volume} {12}},\ \bibinfo {pages} {570} (\bibinfo {year} {1957})},\
  \Eprint {http://arxiv.org/abs/https://doi.org/10.1143/JPSJ.12.570}
  {https://doi.org/10.1143/JPSJ.12.570} \BibitemShut {NoStop}%
\bibitem [{\citenamefont {Kubo}\ \emph {et~al.}(1957)\citenamefont {Kubo},
  \citenamefont {Yokota},\ and\ \citenamefont {Nakajima}}]{kubo2}%
  \BibitemOpen
  \bibfield  {author} {\bibinfo {author} {\bibfnamefont {R.}~\bibnamefont
  {Kubo}}, \bibinfo {author} {\bibfnamefont {M.}~\bibnamefont {Yokota}}, \ and\
  \bibinfo {author} {\bibfnamefont {S.}~\bibnamefont {Nakajima}},\ }\href
  {\doibase 10.1143/JPSJ.12.1203} {\bibfield  {journal} {\bibinfo  {journal}
  {Journal of the Physical Society of Japan}\ }\textbf {\bibinfo {volume}
  {12}},\ \bibinfo {pages} {1203} (\bibinfo {year} {1957})},\ \Eprint
  {http://arxiv.org/abs/https://doi.org/10.1143/JPSJ.12.1203}
  {https://doi.org/10.1143/JPSJ.12.1203} \BibitemShut {NoStop}%
\bibitem [{\citenamefont {Thouless}\ \emph {et~al.}(1982)\citenamefont
  {Thouless}, \citenamefont {Kohmoto}, \citenamefont {Nightingale},\ and\
  \citenamefont {den Nijs}}]{TKNN}%
  \BibitemOpen
  \bibfield  {author} {\bibinfo {author} {\bibfnamefont {D.~J.}\ \bibnamefont
  {Thouless}}, \bibinfo {author} {\bibfnamefont {M.}~\bibnamefont {Kohmoto}},
  \bibinfo {author} {\bibfnamefont {M.~P.}\ \bibnamefont {Nightingale}}, \ and\
  \bibinfo {author} {\bibfnamefont {M.}~\bibnamefont {den Nijs}},\ }\href
  {\doibase 10.1103/PhysRevLett.49.405} {\bibfield  {journal} {\bibinfo
  {journal} {Phys. Rev. Lett.}\ }\textbf {\bibinfo {volume} {49}},\ \bibinfo
  {pages} {405} (\bibinfo {year} {1982})}\BibitemShut {NoStop}%
\bibitem [{Note2()}]{Note2}%
  \BibitemOpen
  \bibinfo {note} {The bandwidth ($9t$) is estimated to be $30meV$ at $3^{\circ
  }$ twist angle \cite {Wang:2020aa}. In other words, the Zeeman splitting
  ($g\mu _BH\sim 7meV$) is about $2t$.}\BibitemShut {Stop}%
\bibitem [{Note3()}]{Note3}%
  \BibitemOpen
  \bibinfo {note} {Current integrated over the upper junction is just opposite,
  because the net current of the entire sample is zero.}\BibitemShut {Stop}%
\end{thebibliography}%

\end{document}